\documentstyle[prl,aps,psfig]{revtex}
\begin{document}
\draft
\twocolumn[\hsize\textwidth\columnwidth\hsize
           \csname @twocolumnfalse\endcsname

\title{Alpha cluster condensation in $^{12}$C and $^{16}$O}
\author{A.~Tohsaki,}
\address{Department of Fine Materials Engineering, Shinshu 
        University, Ueda 386-8567, Japan} 
\author{H.~Horiuchi,} 
\address{Department of Physics, Kyoto University,
       Kyoto 606-8502, Japan}
\author{P.~Schuck,} 
\address{Institut de Physique Nucl\'eaire, 91406 Orsay Cedex, France}
\author{and G.~R\"opke}
\address{FB Physik, Universit\"at Rostock, D-18051 Rostock, 
       Germany}
\date{\today}
\maketitle                              
\begin{abstract}
A new $\alpha$-cluster wave function is proposed which is of the 
$\alpha$-particle condensate type. Applications to $^{12}$C and 
$^{16}$O show that states of low density close to the 3 resp. 4 
$\alpha$-particle threshold in both nuclei are possibly of this kind. 
It is conjectured that all self-conjugate 4$n$ nuclei may show 
similar features.

\end{abstract}

]

Many properties of finite quantum systems such as nuclei, quantum
dots, atomic clusters or ultracould gases in a trap are fairly well
described within a mean-field approximation neglecting the
correlations between the quasiparticles. However, sometimes
correlations become strong, giving rise to the formation of clusters, 
and they have to be taken into account. An intriguing problem is when 
bosonic clusters as bound states of fermions are produced, and the Bose 
character of the composite clusters competes with the fermionic 
properties of their constituents. As an example, we will discuss the 
relevance of $\alpha$-like four-nucleon correlations in atomic nuclei. 
Special attention will be paid to such correlations which correspond 
to an $\alpha$-type condensate in low-density symmetric nuclear matter, 
similar to the Bose-Einstein condensation observed for finite numbers 
of bosonic atoms such as Rb or Na in traps.

It is a well known fact that in light nuclei many states are of the
cluster type\ \cite{cluster,Brink,Bertsch,supple}.  In the case of
cluster states of stable nuclei where we have only very few excess
nucleons in addition to the clusters, they are all located close to or
above the threshold energy of breakup into constituent clusters.  This
fact which is known as the threshold rule\ \cite{ikeho} means that the
inter-cluster binding is weak in cluster states. The threshold rule
can be considered as a necessary condition for the formation of the
cluster structure, because if the inter-cluster binding is strong the
clusters overlap strongly and the clusters will loose their
identities.

For example one of the fundamental questions of the cluster model is 
what kind of $\alpha$-particle cluster states can be expected to exist 
around the threshold energy $E^{\rm thr}_{n \alpha} = n E_{\alpha}$ of
$n\alpha$ breakup in self-conjugate  
$4n$ nuclei.  One possible answer to this question, which is strongly 
under debate, is the existence of the cluster state of a linear 
$n\alpha$ chain structure.  The idea of the linear $\alpha$  chain 
state, originally due to Morinaga\ \cite{morinaga}, is so fascinating 
that recently the formation of linear 6$\alpha$ chain states in 
$^{24}$Mg was studied extensively by experiments and also 
theoretically\ \cite{fulton}.  The possibility of the linear 
3$\alpha$ chain state in $^{12}$C, which is the simplest linear 
$\alpha$ chain state was studied in detail by many authors 
solving the 3$\alpha$ problem microscopically\ \cite{supple}.  However, 
these three-body studies all showed that the 3$\alpha$-cluster states 
around the 3$\alpha$ threshold energy $E^{\rm thr}_{3 \alpha}$ do not
have a linear chain  
structure. For example it was found that the calculated second $0^+$ 
state in $^{12}$C, which corresponds to the observed second $0^+$ state 
located at 0.39 MeV above the 3$\alpha$ threshold energy, has a 
structure where $\alpha$-clusters interact predominantly in relative 
S-waves.  Thus it was concluded that the cluster state near 
$E^{\rm thr}_{3 \alpha}$ has not a linear chain structure but rather 
an $\alpha$-particle gas-like structure.   

On the other hand there have been recent theoretical investigations
on the posibility of $\alpha$-particle condensation in low density
nuclear matter\ \cite{roepschu,roepschk}. In \cite{roepschu} R\"opke 
et al. made a variational ansatz for the solution of the in-medium 
4-body equation. In \cite{roepschk} Beyer et al. solved the Faddeev-
Yakubovsky equations for an alpha-like cluster in nuclear matter. 
The outcome of these studies was that such $\alpha$-condensation can 
occur only in the low-density region below a fifth of the saturation
value.  At higher densities rather a state of ordinary  
$p$-$n$, $n$-$n$, or $p$-$p$ Cooper pairing will prevail.  In view of 
these results it may be a tempting idea that in finite self-conjugate 
$4n$ nuclei one could expect the existence of excited states of 
dilute density composed of a weakly interacting gas of 
$\alpha$-particles.  Since the $\alpha$-cluster is a Bose particle, 
such states could approximately be considered as an $n\alpha$ cluster 
condensed state and eventually excitations thereof.

The purpose of this paper is to report on our study which not only
confirms that indeed the second $0^+$ state in $^{12}$C could be 
considered as such a condensed state but that in addition 
also in $^{16}$O such a state close to the threshold possibly exists.
We will therefore then conjecture that the existence of such 
$\alpha$-condensed states might be a general feature in $N$=$Z$ nuclei.

For the purpose of our study we write down a new type of $\alpha$-cluster
wave function describing an $\alpha$-particle Bose condensed state: 
\begin{equation}
  |\Phi_{n\alpha} \rangle = (C_\alpha^\dagger)^n |{\rm vac} \rangle 
\end{equation}
where the $\alpha$-particle creation operator is given by
\begin{eqnarray}
  C_\alpha^\dagger & = & \int d^3 R\  
     e^{-{\bf R}^2/R_0^2}\ \int d^3 r_1  \cdots d^3 r_4 \nonumber \\ 
     & \times & \varphi_{0s}({\bf r}_1 - {\bf R}) 
     a_{\sigma_1 \tau_1}^\dagger ({\bf r}_1) 
     \cdots \varphi_{0s}({\bf r}_4 - {\bf R}) 
     a_{\sigma_4 \tau_4}^\dagger ({\bf r}_4)
\end{eqnarray}
with $\varphi_{0s}({\bf r}) = (1/(\pi b^2))^{3/4} 
e^{-{\bf r}^2/(2 b^2)}$ and $a_{\sigma \tau}^\dagger({\bf r})$ being
the creation  
operator of a nucleon with spin-isospin $\sigma \tau$ at the spatial 
point ${\bf r}$.  The total $n \alpha$ wave function therefore can be 
written as 
\begin{eqnarray}
&& \langle {\bf r}_1 \sigma_1 \tau_1, \cdots {\bf r}_{4n} \sigma_{4n} 
 \tau_{4n} |\Phi_{n\alpha} \rangle \nonumber \\ && \propto  
  {\cal A}\{ e^{-\frac{2}{B^2} ({\bf X}_1^2 + \cdots + {\bf X}_n^2)}\  
  \phi(\alpha_1) \cdots \phi(\alpha_n) \},
\end{eqnarray}
where $B = (b^2 + 2R_0^2)^{1/2}$ and 
${\bf X}_i = (1/4) \sum_n {\bf r}_{in}$ is the center-of-mass 
coordinate of the $i$-th $\alpha$-cluster $\alpha_i$.  The internal 
wave function of the $\alpha$-cluster $\alpha_i$ is 
$\phi(\alpha_i) \propto \exp [ -(1/8 b^2) \sum_{m>n}^4 
({\bf r}_{im} - {\bf r}_{in})^2 ]$.  The wave function of Eq.(3) is 
totally antisymmerized by the operator ${\cal A}$.  It is to be noted 
that the wave function of Eqs.(1,3) expresses the state where
$n\alpha$-clusters occupy the same $0s$ harmonic oscillator orbit  
$\exp [-\frac{2}{B^2} {\bf X}^2 ]$ with $B$ an indepedent variational 
width parameter.  For example if $B$ is of the size of the whole 
nucleus whereas $b$ remains more or less at the free $\alpha$-particle 
value (a situation encountered below), then the wave function (3) 
describes an $n\alpha$ cluster condensed state in the macroscopic 
limit $n \rightarrow \infty$.  For finite systems we know from the 
pairing case that such a wave function still can more or less reflect 
Bose condensation properties.  Of course the total center-of-mass 
motion can and must be separated out of the wave function of Eq.(1) 
for finite systems.  In the limiting case of $B = b$ 
(i.e. $R_0$ = 0), Eq.(3) describes a Slater determinant of harmonic 
oscillator wave functions. It may also be worth mentioning that for $B 
\neq 0$ the wave function (1,3) is different from the cluster state
proposed by Brink \cite{Brink}.

The state $|\Phi_{n\alpha} \rangle$ has spin-parity $0^+$.  In the
limit of $R_0=0$, the normalized wave function $|\Phi_{n\alpha}^{\rm
  N} \rangle = |\Phi_{n\alpha} \rangle$ $/\sqrt{\langle
  \Phi_{n\alpha}|\Phi_{n\alpha} \rangle}$ is identical to a harmonic
oscillator shell model wave function with the oscillator parameter
$b$. For $n=3$, it is identical to the $p$-shell wave function
$|(0s)^4 (0p)^8, [444]\ 0^+ \rangle$, and for $n=4$, it is identical
to the double closed shell wave function, $|(0s)^4 (0p)^{12}, 0^+
\rangle$. This is easily proved by noticing that these limit wave
functions for $n$ =3 and 4 have maximum spatial symmetry [444] and
[4444], respectively.  Only $^8$Be has an $\alpha$-particle structure
in its ground state. Heavier $n\alpha$ nuclei collapse to the dense
state in their ground state but the individual $\alpha$'s may reappear
when these nuclei are dilated i.e. excited.

We calculated the energy surfaces in the two parameter space,
$R_0$ and $b$, 
$E_{n\alpha}(R_0, b) = \langle \Phi_{n\alpha}^{\rm N}(R_0, b)|{\hat H} 
|\Phi_{n\alpha}^{\rm N}(R_0, b) \rangle$, for $n$ = 3 and 4.
The Hamiltonian ${\hat H}$ consists of the kinetic energy, the 
Coulomb energy, and the effective nuclear force named F1 which was 
proposed by one of the authors  and contains a finite range 
three-nucleon force in addition to the finite range 
two-nucleon force\ \cite{tohsaki}. This force reproduces reasonably 
well the binding energy and radius of the $\alpha$-particle, the 
$\alpha$-$\alpha$ phase shifts of various partial waves, and the 
binding energy and density of nuclear matter.  As we will see below 
this force also gives good results for binding energies and radii of 
$^{12}$C and $^{16}$O.

In Fig.1 we give the contour maps of the energy surfaces 
$E_{n\alpha}(R_0,b)$ for $^{12}$C and $^{16}$O.  The qualitative 
features of both surfaces are similar. They show a valley running from 
the outer region with large $R_0 > 11$ fm and 
$b \approx b_\alpha$ = 1.44 fm to the inner region with small 
$R_0$ and $b > b_\alpha$, where $b_\alpha$ is the oscillator parameter 
of the free  $\alpha$-particle. The valleys have a saddle point at 
$R_0 \approx 10$ fm for $n$ = 3 and at $R_0 \approx 10.6$ fm 
for $n$ = 4.  Beyond the saddle point, 
$E_{n\alpha}(R_0, b_\alpha) \approx E^{\rm thr}_{n \alpha} = n
E_\alpha$, where $E_\alpha$ =  
- 27.5 MeV is the theoretical binding energy of the free $\alpha$ 
particle by the present F1 force in Hartree-Fock approximation.
Therefore we have  $3 E_\alpha$ = -82.5 MeV and $4 E_\alpha$ = -110 
MeV. In the asymptotic region the average inter-$\alpha$ distance is 
large and the kinetic energy of the center-of-mass motion of an 
$\alpha$-cluster ($3\hbar^2/4mB^2$) is very small which leads to 
$E_{n\alpha}(R_0, b_\alpha)$  being more or less equal to 
$E^{\rm thr}_{n \alpha}$. The height of the 
saddle point measured from the theoretical threshold energy is about 
1.4 MeV for $n$ = 3 and 2.2 MeV for $n$ = 4. The appearance of the 
saddle point is due to the increase of the Coulomb energy and kinetic 
energy towards the inward direction which is not yet compensated by 
the gain in potential energy around the saddle point region. This 
saddle point will help to stabilize the possible $\alpha$ condensed 
state around $E^{\rm thr}_{n \alpha}$.
The minimum of the 
energy surface is located at $R_0 \approx$ 2 fm for $n$ = 3 and at 
$R_0 \approx$ 1 fm for $n$ = 4.  Since $R_0 = 0$ means the shell model 
limit, we thus see that the wave function even at the energy minimum 
point deviates from the shell model limit and shows rather strong 
$\alpha$-particle correlations.  The gain in energy from the shell 
model limit is 10.3 MeV for $^{12}$C and 4.7 MeV for $^{16}$O.  Before 
comparing numbers with experiments we have to make a quantum 
mechanical calculation.  This will be achieved via a standard 
Hill-Wheeler ansatz taking $R_0$ and $b$ as the Hill-Wheeler 
coordinates.  However, in order to reduce the complexity of the 
calculation and because the valleys run essentially parallel to the 
$R_0$ axis at $b=b_\alpha$ we take $b = b_\alpha =$ constant and only 
discretise the $R_0$ variable.  We therefore have
\begin{equation}
|\Psi_{n\alpha,k} \rangle = \sum_j f_k((R_0)_j, b_\alpha) 
|\Phi^N_{n\alpha}((R_0)_j, b_\alpha) \rangle. 
\end{equation}
The normalization of $f_k((R_0, b)_j)$ is so that the $k$-th
eigen-function $|\Psi_{n\alpha,k} \rangle$ is normalized. The adopted 
mesh size of $R_0$ values is typically 0.5 fm.  In order to see the 
character of the obtained wave function $|\Psi_{n\alpha,k} \rangle$, 
we introduce the overlap amplitude 
\begin{equation}
A_{n\alpha,k}(R_0,b) = \langle \Phi_{n\alpha}^N(R_0,b)|\Psi_{n\alpha,k} 
\rangle. 
\end{equation}
From this overlap amplitude we can estimate the relevant values of the 
variational parameter $R_0$ in the different states $k$ as will be 
discussed below.

After outlining the results for the new kind of wave function for 
$^{12}$C and $^{16}$O, we will discuss whether the obtained condensed 
states correspond to the states found in these nuclei.  We first 
consider $^{12}$C, i.e. $n$ = 3, see Table I. The calculated lowest 
two eigenenergies are situated at -85.9 and -82.0 MeV. The lowest 
energy state corresponds to the ground state of $^{12}$C and is only 
slightly lower than the minimum point of the energy surface located at 
-85.5 MeV.   However, the calculated ground state is still above the 
observed binding energy of $^{12}$C which is at -92.16 MeV. An 
increase of mesh points  will certainly lower the energy but, as has 
been discussed by many people, in order to reproduce the observed 
$^{12}$C binding energy satisfactorily we have to extend our 
functional space so as to include the spatial symmetry broken wave 
functions which allow to incorporate the effect of the spin-orbit 
force adequately. The second eigenvalue lies 0.36 MeV above our 
theoretical $3\alpha$ threshold energy, $E^{\rm thr}_{3 \alpha}$ =
-82.5 MeV, and we believe that  
it corresponds to the observed second $0^+$ state of $^{12}$C which
lies 0.5 MeV above $E^{\rm thr}_{3 \alpha}$.  As seen in Table I the
rms radius of  
the obtained wave function $|\Psi_{3\alpha,2} \rangle$ is 4.29 fm 
which is much larger than the one of the ground state which is 2.97 
fm, slightly greater than the experimental value 2.45 fm but in 
agreement with the missing binding of 6.75 MeV.  We thus see that the 
second $0^+$ state corresponds to a very dilute system of average 
density which is only 
about a fifth of the experimental ground state density. 

To characterise the wave function by a typical value of the width 
parameter $R_0$, we consider the overlap amplitude 
$A_{3\alpha,k}(R_0,b_\alpha)$ given in Eq.(6) as a function of $R_0$ 
at fixed $b_\alpha$. Whereas the ground state ($k=1$) wave function 
is almost exhausted by one $\Phi^N_{3\alpha}(R_0,b_\alpha)$ with 
$R_0 \approx 2$ fm which is quite close to the wave function of the 
minimum energy point of the energy surface, the second $0^+_{k=2}$ 
state has the largest overlap amplitude (about 0.87) with 
$R_0 \approx 4.5$ fm. This rather large value implies that the 
distribution of the center-of-mass momenta is rather narrow, in a 
certain approximation to an $\alpha$ condensate in infinite nuclear 
matter where all $\alpha$-clusters populate the same state $P=0$ of 
the center-of-mass momentum. The fact that the calculated $0^+_{k=2}$ 
state is of dilute density is in agreement with nuclear matter 
calculations\ \cite{roepschu,roepschk} where it was shown that a 
condensate of $\alpha$-like particles (quartetting) is possible only 
in matter with density $\rho \le 0.03$ fm$^{-3}$. The average distance 
of the $\alpha$-clusters in the dilute $0^+_{k=2}$ state is in 
agreement with this value for low density nuclear matter, where the 
overlap of the $\alpha$-clusters is small so that the Pauli blocking 
effects are weak.

Let us now discuss the case of $^{16}$O, i.e. $n=4$. The energies of
the lowest observed $0^+$ states are shown in Table II, together with
the corresponding widths. The first excited $0^+_2$ state at 6.06 MeV
is very well known to have $\alpha$-clustering character\ 
\cite{cluster,supple} and is well described by the $^{12}$C + $\alpha$
microscopic cluster model as having the structure where the
$\alpha$-cluster moves in a S state around the $^{12}$C-cluster in its
ground state\ \cite{suzuki} though also other cluster states have been
proposed \cite{Bertsch}. Similarly, the third excited $0^+_4$ state at
12.05 MeV can be described by the same model where the
$\alpha$-cluster moves in a D state around the $^{12}$C-cluster in its
first 2$^+$ excited state\ \cite{suzuki}.  We will exclude these well
understood states from our further discussion. The excited states
$0^+_3$ at 11.26 MeV and $0^+_5$ state at 14.0 MeV observed in
$^{12}$C + $\alpha$ elastic scattering\ \cite{ajzsel} cannot be
described by such a model. Furthermore, they have very large decay
widths, not typical for the other states. These states may be
described by our new wave function as condensed states.

As seen from Table I, the experimental value of the ground state
($0^+_1$) of $^{16}$O at -127.62 MeV is well reproduced by the
calculated energy value for the ground state ($0^+_{k=1}$) at -124.8
MeV. The calculated energy is above the minimum energy of the energy
surface. It is because the $b$ value of the minimum energy point is
fairly larger than $b_\alpha$ and the minimum energy point is not
covered by the adopted mesh points. In order to have a better wave
function for the ground state we, of course, need to include mesh
points around the minimum energy in our generator coordinate
calculation.  When we adopt $b$ = 1.57 fm which is the $b$ value of
the energy minimum point, the generator coordinate calculation gives
-128.0 MeV as the lowest eigen energy.  The rms radius of the
calculated ($0^+_{k=1}$) state is 2.59 fm and is slightly smaller than
the observed ($0^+_1$) rms radius, 2.73 fm, of $^{16}$O.  The second
($0^+_{k=2}$) state of our calculation is bound by 6 MeV below the
theoretical 4$\alpha$ threshold energy. The rms radius of this state
is 3.12 fm and this state has the largest overlap amplitude (about
0.86) with $\Phi_{4 \alpha}^N(R_0,b_\alpha)$ with $R_0 \approx 2.5$
fm. We conjecture that this state corresponds to the observed $0^+_3$
state situated at 3.18 MeV below the observed 4$\alpha$ threshold
energy. Indeed one may argue that there will be some mixing between
the second ($0^+_{k=2}$) state and the $^{12}$C+$\alpha$ state,
bringing theoretical and experimental energies closer together. The
third ($0^+_{k=3}$) state of our calculation is bound by 0.7 MeV below
the theoretical 4$\alpha$ threshold energy.  We think that it may
correspond to the measured $0^+_5$ state situated at 0.44 MeV below
the observed 4$\alpha$ threshold energy. This state has a very large
rms radius of 3.94 fm and has the largest overlap amplitude (about
0.80) with $\Phi_{4 \alpha}^N(R_0,b_\alpha)$ with $R_0 \approx 4.1$
fm. In analogy to the case of $^{12}$C, these values
indicate that this state of dilute density should be considered as
just the 4$\alpha$-cluster condensed state that we expected. One
should point out that the ease with which we get the $0^+$-states
around $E^{\rm thr}_{n \alpha}$ is a strong indication that our wave
function (1,3) grasps the essential physics because otherwise the
threshold states are very difficult to obtain. We also would like to
mention that the present formalism yields very good results for the
groundstate of $^8$Be as well.

In conclusion, our present study thus predicts in $^{12}$C and 
$^{16}$O the existence of near-$n\alpha$-threshold states which are 
the finite system analogues to $\alpha$-cluster condensation in 
infinite matter.  They are characterized by low density states so that 
the $\alpha$-clusters are not strongly overlapping and by an $n$-fold
occupation of their identical S-wave center-of-mass wave function.
Therefore these states are quite similar in structure to the 
Bose-Einstein condensed states of bosonic atoms in magnetic traps 
where all atoms populate the same lowest S-wave quantum orbital. 
Because of the short life-time of the $\alpha$-condensed states, the 
predicted large values for the rms radii may be verified by indirect 
methods.  The measurements of the spectra of the emitted $\alpha$ 
particles should allow to determine the Coulomb barrier which is 
expected to be small for the low density states.  Of particular 
interest would be $\alpha$-$\alpha$ coincidence measurement of 
decaying condensed states.  

We conjecture that such condensed 
$\alpha$-cluster states near the $n\alpha$ threshold may also occur 
in other heavier 4$n$ self-conjugate nuclei.  For example condensed 
6$\alpha$ states of $^{24}$Mg could be deformed and the measurement 
of a reduced moment of inertia over the rigid body value would be a 
strong indication for $\alpha$-particle superfluidity.  The wave 
function we have proposed in this work is very flexible and can 
straightfowardly be adopted for the description of other condensation 
phenomena such as for example ordinary Cooper pairing or a mixture of 
Cooper pair and $\alpha$-particle condensation. 

\normalsize
\rm

\newpage
\onecolumn
\begin{table}[bp]
\begin{center}
\begin{tabular}{cc|cccccc} 
\hline\hline
 &   & $E_k$ & $E_{\rm exp}$  & $E_k-E^{\rm thr}_{n \alpha}$  &
 $(E-E^{\rm thr}_{n \alpha})_{\rm exp}$  & $\sqrt{\langle r^2\rangle}$&
 ${\sqrt{\langle r^2\rangle}}_{\rm exp}$ \\
 & & (MeV) & (MeV) & (MeV) & (MeV) & (fm)  & (fm)  \\\hline
$^{12}{\rm C}$ & $k=1$ & -85.9  & -92.16 $(0_1^+)$ & -3.4 & -7.27 &
2.97 & 2.65 \\ 
 & $k=2$ & -82.0  & -84.51 $(0_2^+)$ & +0.5 & 0.38 & 4.29 &  \\
 & $E^{\rm thr}_{3\alpha}$ & -82.5  & -84.89 &  &  &  &  \\\hline
$^{16}{\rm O}$ & $k=1$ & -124.8& -127.62 $(0_1^+)$ & -14.8
 & -14.44 & 2.59 & 2.73 \\
 & & $(-128.0)^*$  & &$(-18.0)^*$  & & &\\
 & $k=2$ & -116.0  & -116.36 $(0_3^+)$ & -6.0  & -3.18 & 3.16 &  \\
 & $k=3$ & -110.7  & -113.62 $(0_5^+)$ & -0.7  & -0.44 & 3.97 &  \\
 & $E^{\rm thr}_{4\alpha}$ & -110.0  & -113.18 &  &  &  &  \\\hline\hline
\end{tabular}
\end{center}
\caption{Comparison of the generator coordinate method calculations 
with experimental values. $E^{\rm thr}_{n\alpha} = n E_{\alpha}$
denotes the threshold energy for the decay into $\alpha$-clusters, the 
values marked by * correspond to a refined mesh, see main text.}
\label{TABLE-I}
\end{table}

\begin{table}[bp]
\begin{center}
\begin{tabular}{c|cc} 
\hline\hline
 & $E_{\rm exc}$ (MeV) & $\Gamma$ (MeV) \\\hline
$0_2^+$ & 6.06  &  \\
$0_3^+$ & 11.26  & 2.6 \\
$0_4^+$ & 12.05  & 1.6$\times 10^{-3}$ \\
$0_5^+$ & 14.0  & 4.8 \\
$0_6^+$ & 14.03  & 2.0$\times 10^{-1}$ \\\hline\hline
\end{tabular}
\end{center}
\caption{Observed excitation energies $E_{\rm exc}$ and widths
  $\Gamma$ of the lowest five  
$0^+$ excited states in $^{16}$O}
\label{TABE-II}
\end{table}

\begin{figure}
\centerline{\psfig{figure=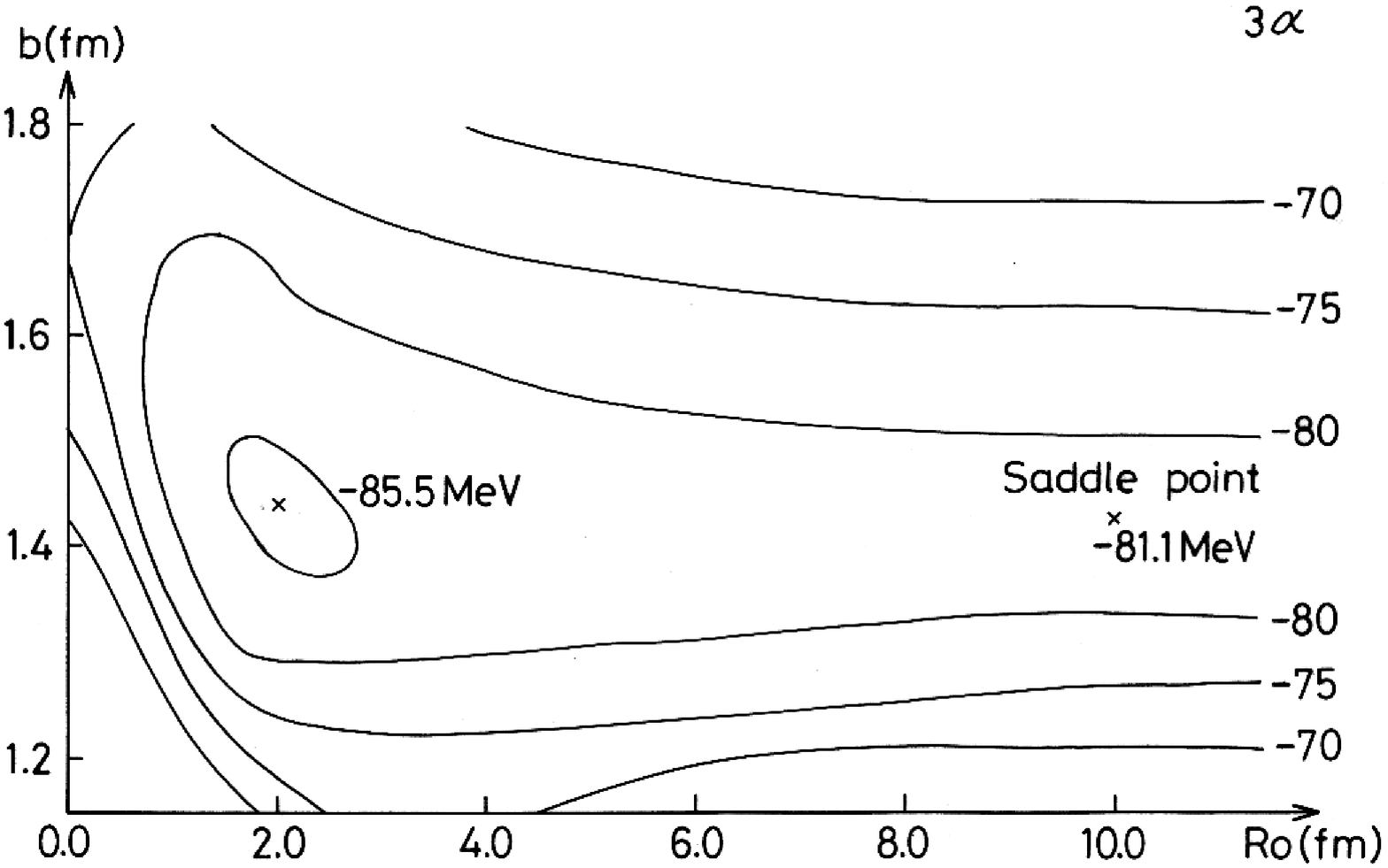,height=9cm,width=13cm,angle=0}}
\vspace{1cm}
\caption{\label{fig1} 
 Contour map of the energy surface $E_{3\alpha}(R_0,b)$ for 
$^{12}$C. Numbers attached to the contour lines are the binding 
energies. }
\end{figure}

\begin{figure}
\centerline{\psfig{figure=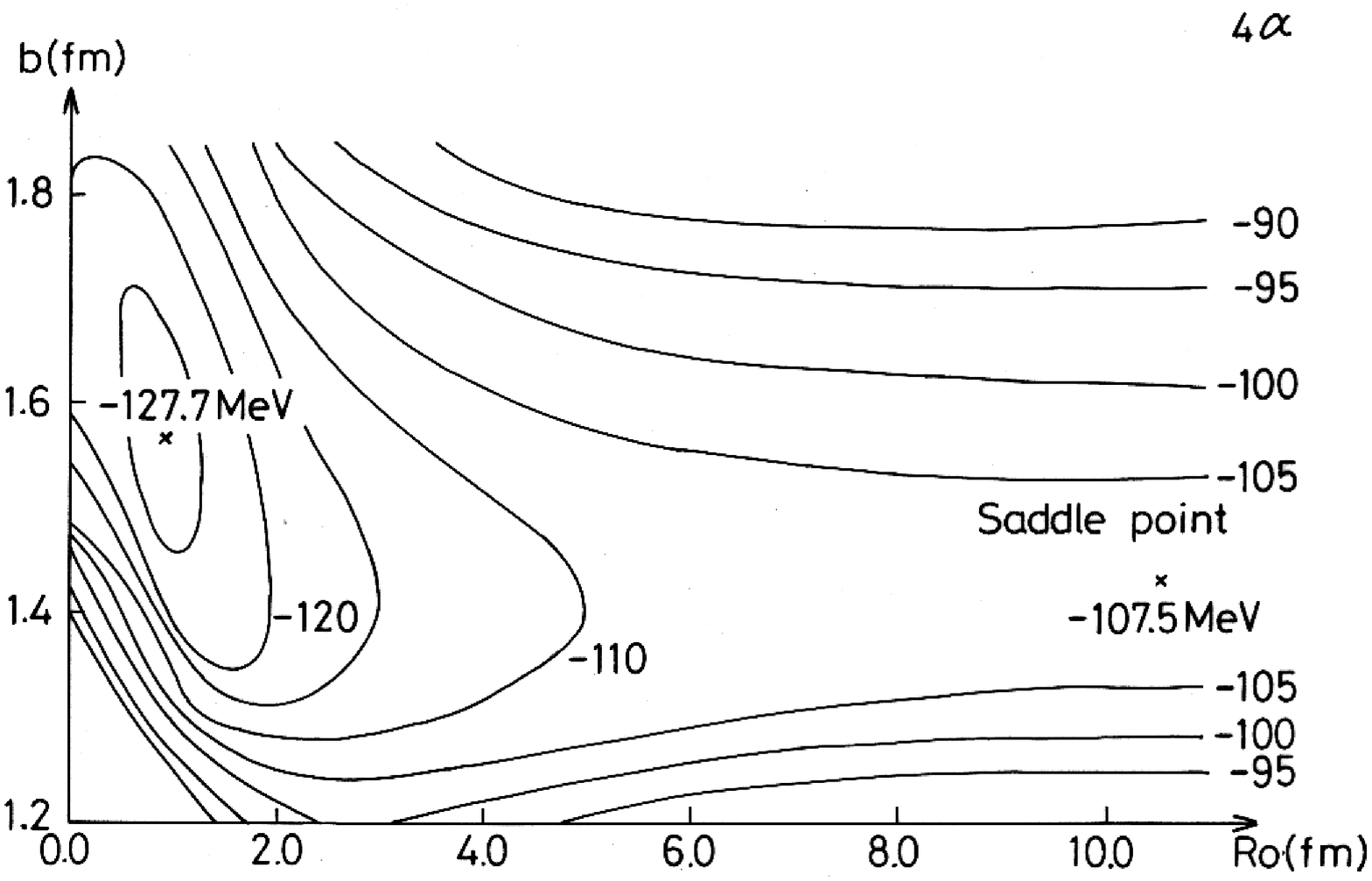,height=9cm,width=13cm,angle=0}}
\vspace{1cm}
\caption{\label{fig1} 
Contour map of the energy surface $E_{4\alpha}(R_0,b)$ for 
$^{16}$O. Numbers attached to the contour lines are the binding 
energies.}
\end{figure}

\end{document}